# Robust Attack Detection Approach for IIoT Using Ensemble Classifier

V. Priya[1], I. Sumaiya Thaseen[1], Thippa Reddy Gadekallu[1], Mohamed K. Aboudaif[2,*] and Emad Abouel Nasr[3]

[1]School of Information Technology and Engineering, Vellore Institute of Technology, Vellore, 632014, India
[2]Advanced Manufacturing Institute, King Saud University, Riyadh, 11421, Saudi Arabia
[3]Industrial Engineering Department, College of Engineering, King Saud University, Riyadh, 11421, Saudi Arabia
*Corresponding Author: Mohamed K. Aboudaif. Email: maboudaif@ksu.edu.sa

**Abstract:** Generally, the risks associated with malicious threats are increasing for the Internet of Things (IoT) and its related applications due to dependency on the Internet and the minimal resource availability of IoT devices. Thus, anomaly-based intrusion detection models for IoT networks are vital. Distinct detection methodologies need to be developed for the Industrial Internet of Things (IIoT) network as threat detection is a significant expectation of stakeholders. Machine learning approaches are considered to be evolving techniques that learn with experience, and such approaches have resulted in superior performance in various applications, such as pattern recognition, outlier analysis, and speech recognition. Traditional techniques and tools are not adequate to secure IIoT networks due to the use of various protocols in industrial systems and restricted possibilities of upgradation. In this paper, the objective is to develop a two-phase anomaly detection model to enhance the reliability of an IIoT network. In the first phase, SVM and Naïve Bayes, are integrated using an ensemble blending technique. K-fold cross-validation is performed while training the data with different training and testing ratios to obtain optimized training and test sets. Ensemble blending uses a random forest technique to predict class labels. An Artificial Neural Network (ANN) classifier that uses the Adam optimizer to achieve better accuracy is also used for prediction. In the second phase, both the ANN and random forest results are fed to the model's classification unit, and the highest accuracy value is considered the final result. The proposed model is tested on standard IoT attack datasets, such as WUSTL_IIOT-2018, N_BaIoT, and Bot_IoT. The highest accuracy obtained is 99%. A comparative analysis of the proposed model using state-of-the-art ensemble techniques is performed to demonstrate the superiority of the results. The results also demonstrate that the proposed model outperforms traditional techniques and thus improves the reliability of an IIoT network.

## Abbreviations

ACO        Ant Colony Optimization
ANN        Artificial Neural Network

| | |
|---|---|
| BPN | Back Propagation Network |
| CNN | Convolutional Neural Network |
| GRU | Gated Recurrent Unit |
| LSTM | Long Short Term Memory Networks |
| RNN | Recurrent Neural Network |
| CPS | Cyber Physical Systems |
| DT | Decision Tree |
| FNT | Flexible Neural Tree |
| GA | Genetic Algorithm |
| IoT | Internet of Things |
| IIoT | Industrial Internet of Things |
| KNN | K-Nearest Neighbor |
| KPCA | Kernel Principal Component Analysis |
| NB | Naïve Bayes |
| PCA | Principal Component Analysis |
| RF | Random Forest |
| SA | Simulated Annealing |
| SVM | Support Vector Machine |

## 1 Introduction

Currently, the number of IoT devices and connected devices is estimated to be more than 15 billion, and up to 50 billion connected IoT devices are expected by 2022. Development of huge numbers of IoT devices combined with the pressure to deliver IoT devices to market in a timely and competitive manner has increased attention on privacy and security issues. Advances in the IoT and Cyber Physical System (CPS) domains has stimulated creation of Cyber-Physical Manufacturing Systems (CPMS). With the continuous development of CPMSs, significant security concerns have been raised in relation to the Industrial IoT (IIoT), which is characterized by real-time monitoring, automated systems, smart connections, and collaborative machines [1]. Identifying IIoT threats and developing defense strategies is required because the complete internet could be paralyzed if a single component and/or communication channel in an IIoT-based system is compromised.

The four-layered architecture of the IIoT is shown in Fig. 1. The first layer is the edge layer, which contains the IIoT devices, and the second layer, the aggregation layer, consists of connected devices. The third layer is the network layer. The fourth layer is the cloud layer, which performs analytics, reporting, and planning based on data captured from the IIoT devices. As shown in Fig. 1 (edge layer), IIoT devices will be distributed in various environments, including remote locations where routine maintenance is not feasible. Furthermore, the control logic on IIoT devices cannot be determined in the destination environment. IIoT devices are vulnerable to various types of attacks, such as DDoS, DoS, tampering, spoofing, privilege escalation, and IoT botnet attacks [2].

Cisco analyzed a survey [3] that identified Trojan as the most common type of malware deployed to access users and an organization's computers. Security is a significant challenge that has to be addressed sensibly. As shown in Fig. 2, the global cybersecurity market has increased due to increasing threats and attacks, and, by 2023, it is expected that the market will increase exponentially. Despite measures implemented to maintain a secure environment, attacks can occur [4]. Building a network that is immune to all types of attack is not possible. Therefore, to realize a trust-based IoT network, developing ways to preventing or mitigate attacks is very important.

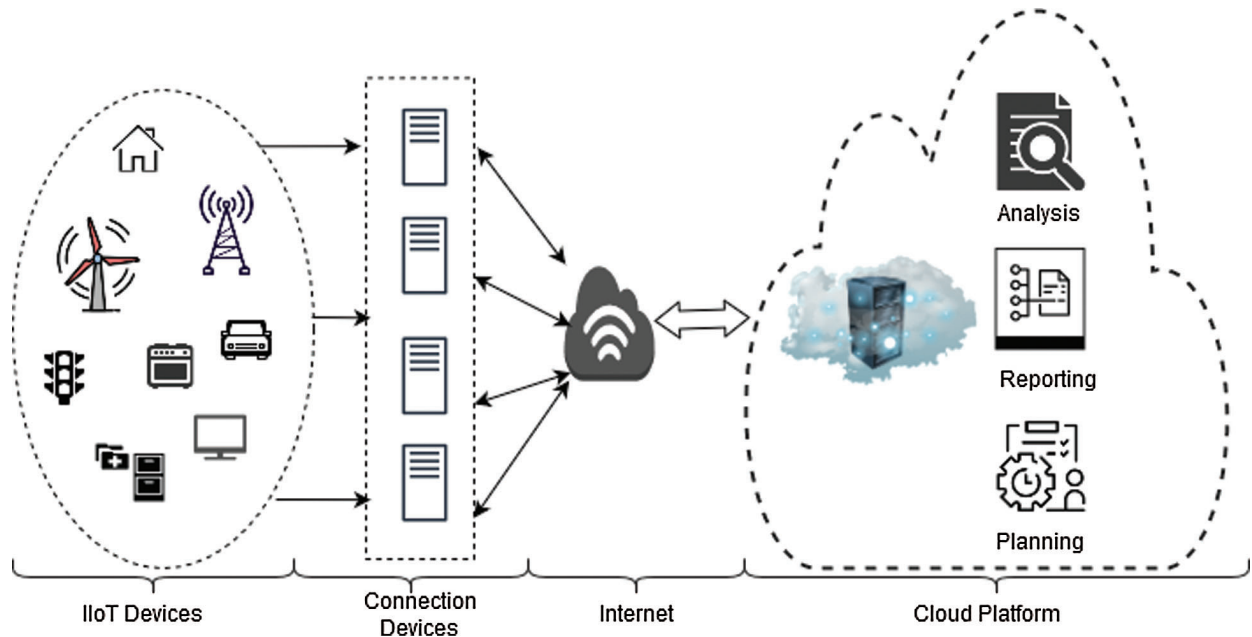

**Figure 1:** IIoT Architecture

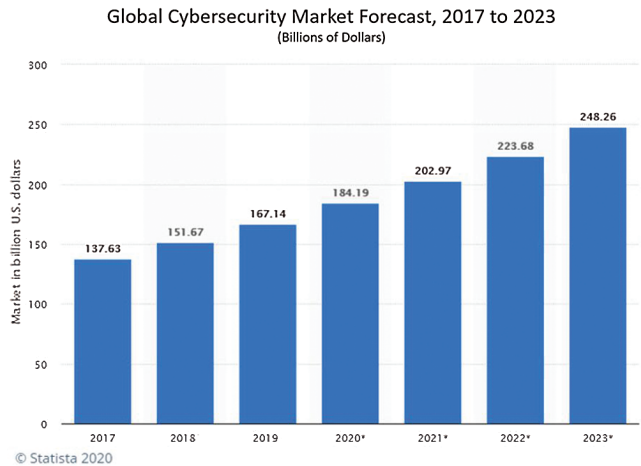

**Figure 2:** Global cybersecurity market projected to 2023

Security solutions use antivirus software and intermediate boxes, such as Intrusion Detection Systems (IDS) and firewalls. A firewall controls inbound and outbound traffic at the network endpoints based on the source and destination addresses. However, firewalls require knowledge of the host and are limited by the amount of state available. IDSs are security monitoring tools. They analyze network traffic and scan the system for malicious activities. In addition, IDSs notify the system administrator when a malicious incident is detected. Misuse, anomaly, and hybrid detection mechanisms are widely used in IDSs. With misuse identification, unknown attacks are detected by knowledge rules. In anomaly detection, attacker behavior is compared to normal behavior based on a hypothesis. Hybrid techniques integrate misuse and anomaly detection mechanisms.

Various machine learning approaches have been developed for anomaly detection in the IoT. Methods based on machine learning have proven to be effective for identifying anomalous events in the network traffic flow. Machine learning strategies can be classified as supervised and unsupervised. Unsupervised learning does not require labelled data. However, with supervised learning, the algorithm is train on labelled samples; i.e., the process includes a function whereby samples are mapped to class labels. In the testing phase, the class for the unpredicted samples is determined according to the function. Widely used machine learning techniques include Naïve Bayes, SVM, KNN, and decision trees [5]. Convolutional Neural Networks (CNN) [6,7] are also employed in machine learning. There are many ensemble techniques, such as random forest [8], Bending, AdaBoost, and stacking. However, there is no universal approach that can work equally well on all datasets [9].

In this paper, a unified two-phase intrusion detection model is developed using an ensemble machine learning approach called blending that integrates SVM, NB, and DT in the first phase and a random forest classifier is used for prediction. In addition, the results of an Artificial Neural Network (ANN) classifier are integrated with those of the random forest to obtain the best prediction. A contingent analysis is performed by evaluating the integrated model against the WUSTL_IIOT-2018, N_BaIoT and Bot_IoT datasets. In this analysis, accuracy, precision, F-Score, and recall are measured.

The primary contributions of this study are as follows.

- Several existing studies on intrusion detection in the IoT are examined. The investigation focuses on the performance of the algorithms used to develop an attack identification approach.
- Base and ensemble machine learning techniques are integrated to construct a robust approach for anomaly detection.
- Accuracy and other performance metrics on various benchmark IoT datasets are analyzed.

The remainder of this paper is organized as follows. A brief review of related work is presented in Section 2. Section 3 addresses the proposed IIoT attack identification model. The results and performance analysis of the proposed model on various datasets are discussed in Section 4. Conclusions and suggestions for future work are provided in Section 5.

## 2 Literature Survey

Machine learning approaches are known to provide optimal intrusion detection solutions. Compared to other methods, machine learning approaches provide better results because they can be applied to various types of datasets and can analyze real-time data. In a previous study, a trust model was constructed for machine-to-machine communication using various machine learning approaches, such as logistic regression, NB, DT, KNN and RF [4]. A comparative study has been performed to identify the best approach [5]. That study investigated various techniques, i.e., Naïve Bayes, an SVM, and decision trees. This approach provides accurate information regarding anomalous behaviors and can also analyze the source of the intrusion or the main issue. Typically, these problems are detected based on data patterns, which is time-consuming for human analysts. In this study, large data sets were evaluated, which is labor-intensive and time-consuming with conventional approaches.

Deep learning approaches, such as CNN, CNN-LSTM, CNN-RNN and CNN-GRU, have also been used to identify intrusions [6]. These approaches have proven to be more accurate; however, due to the complex architecture, a high computational cost is incurred during training. To increase accuracy, an ANN model that used a wrapper method for feature selection was constructed [8]. The proposed ANN model was compared to an SVM, and the comparison shows that the proposed model yielded more accurate results. Simulated annealing with an SVM is a hybrid approach that has been applied to network intrusion [10]. This approach proved to be significantly more accurate than an SVM alone.

The limitation of this approach that more false positives are generated compared to other methods, such as BPN. The deployment of machine learning approaches in cybersecurity has been analyzed [11]. In addition, multiple classifier techniques have been studied [12]. In that study, the misuse detection model is combined with anomaly detection. A decision tree was used in the anomaly detection module. This approach proved to be effective as it minimized the number of false positives, and the rate of detection was improved.

Bhattacharya et al. [13] constructed a network intrusion detection model that used an integrated PCA-Firefly-based XGBoost approach. In that study, PCA is applied to reduce dimensionality, and XGBoost, which is an advanced ensemble method, was used to predict the classification. Another study, proposed an intrusion identification model using hybrid PCA-GWO for IoMT [14]. The proposed model resulted in better accuracy and decreased the time complexity by 32% for faster alert generation. Rupa et al. [15] analyzed various classifiers, such as LinearSVC, logistic regression, MultinomialNB, and random forest, and developed a computational system that could classify cyber-crime offences. The results demonstrated that logistic regression outperformed superior all other analyzed classifiers.

Significant machine learning algorithms deployed on various benchmark datasets are listed in Tab. 1. The DT, NB, ANN, and RF classifiers obtained maximum accuracy of 99% on at least one benchmark dataset. However, maximum accuracy for the SVM was 96% due to its known generalization issue. These results led to the selection of these classifiers for the proposed integrated anomaly detection model for IoT because a blend of these classifiers could result in increased accuracy and reduced error rates.

## 3 Proposed Methodology

The proposed methodology (Fig. 3) is an efficient method that provides a trust-based attack identification model for a network. Initially, the datasets, i.e., WUSTL_IIOT-2018, N_BaIoT and Bot_IoT, are normalized. In the initial stage, the values are fitted between 0 and 1 using label encoding to avoid overfitting. Another level data preprocessing is performed using the Standard Scaler to eliminate null and redundant data. The Standard Scaler arranges the data in a standard normal distribution. In the next step, the data are divided with different cross-validation ratios, e.g., 60:40, 70:30, and 80:20. It was observed that an 80:20 ratio results in better accuracy at the first level of deployment. This model ensures that all observations from the dataset have a fair chance of appearing in the training and test data. A two-level of classification is deployed in the proposed model. In the first level, SVM, Naïve Bayes, and a decision tree are integrated as a blended ensemble, and the output is a new training set that is sent to a random forest classifier. In addition, an ANN classifier is deployed on the data using softmax as the activation function. Here, the Adam optimizer is used to improve accuracy. In the second level, both the ANN and random forest results are sent to the classification unit, and the most accurate result is considered the final predicted test result. The pseudocode of the proposed model is shown in Fig. 4.

2462 CMC, 2021, vol.66, no.3Table 1: Evaluation of significant machine learning techniques in cybersecurity

| Technique | Dataset | Ref. No. | Domain | Accuracy (%) | Precision (%) | Recall (%) |
|---|---|---|---|---|---|---|
| Naive Bayes | DARPA | [16] | Misuse-Based | 99.90 | 99.04 | 99.50 |
| | NSL-KDD | [17] | | 81.66 | – | – |
| | KDD CUP99 | [18] | Signature-Based | 99.72 | – | 100 |
| ANN | DARPA | [19] | Misuse-Based | 99.82 | – | – |
| | NSL-KDD | [20] | Anomaly-Based | 94.50 | – | – |
| | KDD CUP99 | [21] | | 62.90 | – | – |
| SVM | DARPA | [22] | Anomaly-Based | 95.11 | – | – |
| | NSL-KDD | [23] | Anomaly-Based | 89.70 | – | – |
| | KDD CUP99 | [24] | Hybrid-Based | 96.08 | – | – |
| Decision Tree | KDD | [19] | Misuse | 99.96 | | |
| | NSL-KDD | [25] | Hybrid | 93.40 | | |
| | KDD CUP99 | [26] | Hybrid | 92.87 | 99.90 | |
| Random Forest | KDD | [27] | Anomaly | 99.95 | – | 99.95 |
| | NSL-KDD | [28,29] | Anomaly-Based | 96.30 | 99.80 | |
| | | | Hybrid-Based | 75.30 | 81.40 | 75.30 |

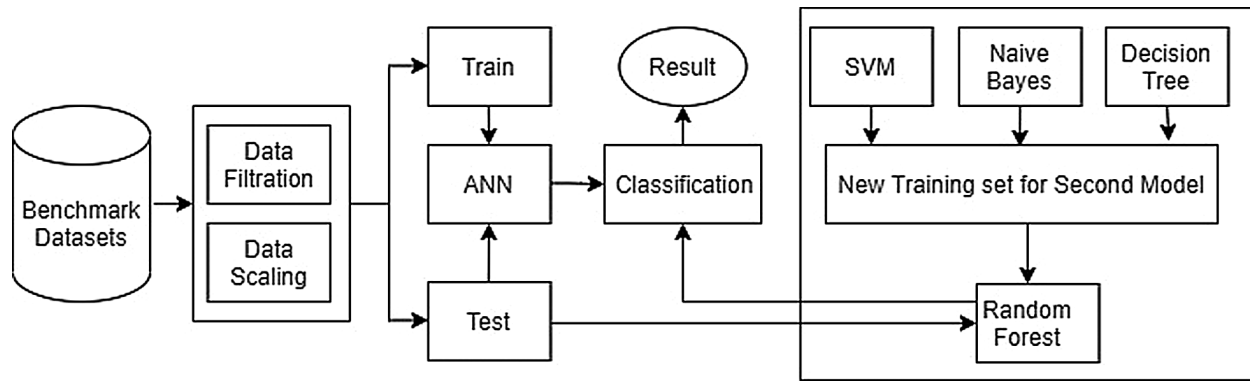

Figure 3: Proposed IIoT attack identification model

## 4 Experimental Analysis

### 4.1 Dataset Description

The first dataset used is the WUSTL_IIOT_2018 dataset for ICS (SCADA) Cybersecurity [30]. Real-world industrial systems are closely emulated, and cyber attacks are generated and captured. The different attacks generated in the testbed are listed in Tab. 2. The dataset contains 93.93% normal traffic and 6.07% abnormal traffic. Initially, the dataset has 25 features. However, based on an analysis, six features are selected, as shown in Tab. 3. After the data are cleaned to eliminate null and redundant data, a new column is introduced as "Target" wherein normal traffic is represented as "0" and attack traffic is represented as "1".

Table 2: Attacks generated in WUSTL_IIOT_2018 dataset

| Attack | Description |
| --- | --- |
| Port Scanner | The attack is difficult to identify because the TCP connection is not established completely. |
| Address Scan | Network and Modbus addresses are scanned. The unique address of the Modbus server is utilized for future attacks. |
| Device Identification | The Modbus slave IDs on the system are enumerated and extra information is collected. |
| Device Identification (Aggressive) | The scan is performed in forceful mode to gather supplementary information about slave IDs. |
| Exploit | The coil values of SCADA devices are read. Coil values indicate the ON/OFF status of a device. |

Table 3: Features selected in WUSTL_IIOT_2018 dataset

| Features | Description |
| --- | --- |
| Source Port | Source port number |
| Total packets | Count of the total transaction packets |
| Total Bytes | Transaction bytes |
| Source Packets | Source packet sum |
| Destination Packets | Destination packet sum |
| Source Bytes | |

The second dataset used for our analysis is the N_BaIoT dataset [31] that comprises data from nine commercial IoT devices infected by the Bashlite and Mirai botnets. The data is classified as malicicious (10 categories) and benign (1 category). Initially, the datasets had more than 100 features. However, after stream aggregation and deploying statistics, 12 features are used for analysis.

The final dataset is the BoT_IoT [32] generated by the Australian Centre for Cyber Security. This dataset contains both anomalous and normal events. There are six attack categories, i.e., Data exfiltration, Service Scan, DDoS, Keylogging, DoS, and OS attacks.

### 4.2 Pseudocode of the Proposed Model

*Input : Data D = {$a_i, b_i$} where i=1 to n*
*Output : Class Label*

1. Step 1: Employ first-level Classifiers
2. for t <- 1 to 3 do
3. Create the subset $S_m$ filtering.
4. Learn and evaluate three supervised classifiers $h_t$ based on D where $h_0$- SVM, $h_1$- Naïve Bayes and $h_2$- Decision Tree using the subset $S_m$.
5. End for.
6. Step 2 : Create different datasets from D
7. for t <-1 to 3 do
8. Construct a new data set that contains {$a_i, b_i$}, where $a_i = \{h_1(a_i), h_2(a_i), h_3(a_i)\}$
9. Initialize the base classifiers for the class $\omega_k, w_k = 0$
10. Get the maximum performance for the class $w_k$

$$B_k = \max_{m=1}^{N} \left\{ \max_{n=1}^{N} \left\{ R_{C_{N,N}}^k \right\} \right\} \tag{1}$$

11. Get the maximum performance for the class $w_k$,

$$T_k = B_k * \beta \tag{2}$$

12. Step 3: Learn the next level classifier Random Forest ($h_4(a)$).
13. Step 4: Learn a new classifier ANN ($h_5(a)$) based on the newly derived data set.
14. Initialize all weights and biases
15. while terminating condition is not obtained {
16. for every training sample X in D{
17. for every input layer unit 'r'{
18. $O_r = I_r$
19. for every hidden or results layer unit 'j' {
20. $I_j = \sum_r w_{rj} O_r + \phi_j;\ O_j = 1 + (1 + e^{-1}j)$ (3)

21. for every unit in the output layer
22. $Err_j = O_j(1 - O_j)(T_j - O_j)$ (4)

23. for ever unit in the hidden layer,
24. $Err_j = O_j(1 - O_j) \sum_k Err_k w_{jk}$ (5)

25. For every weight $w_{jk}$ in the network
26. $\Delta w_{rj} = (l)\ Err_j\ O_j; w_{rj} = w_{rj} + \Delta w_{rj}$
27. for each bias $\theta_j$ in the network {

$\Delta \theta_j = (l)\ Err_j;\ Q_r = Q_r + \Delta Q_i; \}\}$

28. Step 5 : $H(a) = \arg\max \sum_{i=1 to 2} 1(y = h_i(a))$ # the value of 1(∝) is 1 if ∝ is true, % and 0 otherwise.

**Figure 4:** Pseudocode of the Proposed Model

### 4.3 Performance Metrics

● Accuracy: Accuracy measures the correctness of a result. In this case, the correctness of the model's predictions are measured. Accuracy can be expressed as follows.

$$(t_p + t_n)/(t_p + f_p + t_n + f_n) \tag{8}$$

- Precision (P): Precision represents the exactness of a classifier and can be expressed as follows.

$$t_p/(t_p + f_p) \tag{9}$$

- Recall(R): Recall defines the completeness of a classification model Recall can be expressed as follows.

$$t_p/(t_p + f_n) \tag{10}$$

- F1 score: The F1 score measures acccuracy based on precision and recall values. F1 values are calculated as follows.

$$2 * ((P * R)/(P + R))$$

Here, true negatives, true positives, false positives, and false negatives are represented as $t_n$, $t_p$, $f_p$, and $f_n$, respectively.

### 4.4 Results

Performance indicators, such as accuracy, precision, F1 score, and recall are measured to evaluated the proposed model. The various performance metrics obtained using the SVM, NB, and DT classifiers [33,34] on the WUSTL_IIOT_2018 dataset are shown in Fig. 5. Naive Bayes performs poorly with accuracy, precision, recall, and f-score values of 83, 86, 84, and 83, respectively. The SVM and DT classifiers results were similarly; the DT classifier demonstrated maximum accuracy of 96%.

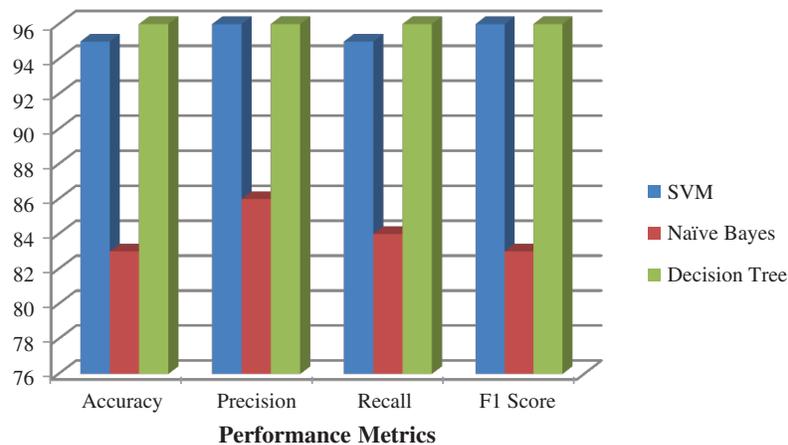

**Figure 5:** Performance metrics obtained on first level classification on WUSTL_IIOT_2018 dataset

The performance metrics of the proposed model on the WUSTL_IIOT_2018 dataset after the second level of classification are shown in Fig. 6. The classification result is obtained by predicting the better of the random forest and ANN classifier results. Here, maximum accuracy of 99% is obtained. Note that the Adam optimizer is deployed for the ANN as it can rapidly converge and has a high variance. Thus, a two-level classification results in the best prediction.

The various performance metrics obtained using the SVM, NB, and DT classifiers on the N_BaIoT dataset are shown in Fig. 7. NB performs poorly with accuracy, precision, recall. and F-scores of 87, 88, 88, and 87, respectively. The SVM returns 95% accuracy, which is better than the NB classifier. The DT classifier outperforms both the SVM and the NB classifiers with maximum accuracy of 98%. The results of the proposed model after the second level of classification are shown in Fig. 8. The result of the blending is used to train new data and send it to the RF classifier. The ANN and RF predictions are merged to derive a new result with an accuracy of 99%.

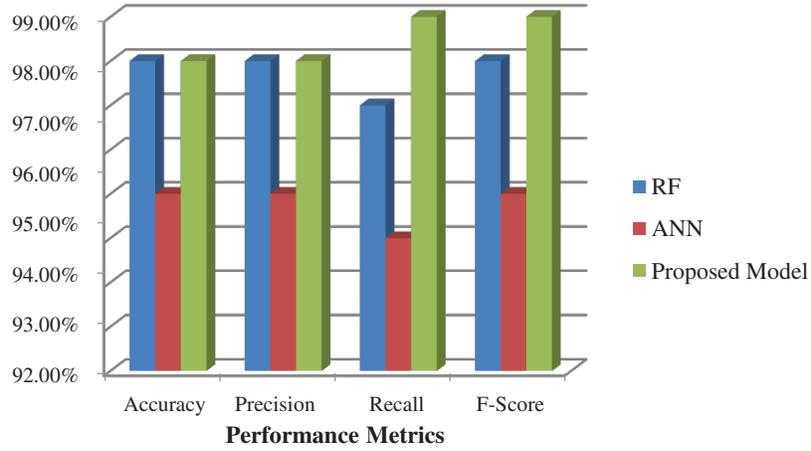

**Figure 6:** Performance metrics of the integrated model using WUSTL_IIOT_2018 dataset

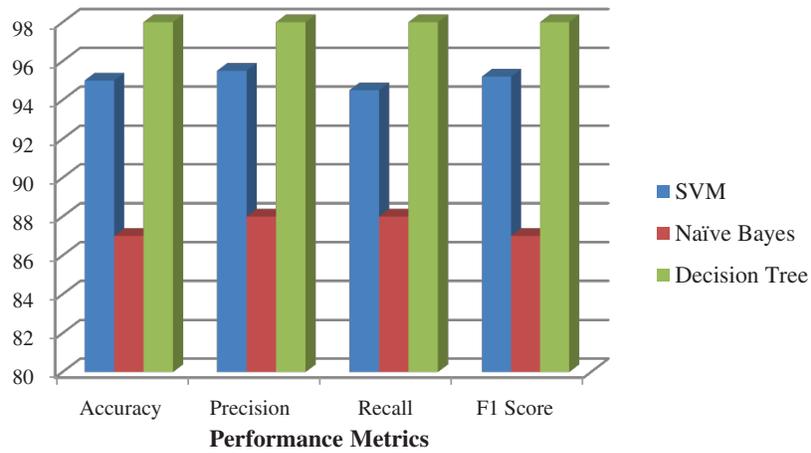

**Figure 7:** Performance metrics obtained on first level classification on N_BaIoT dataset

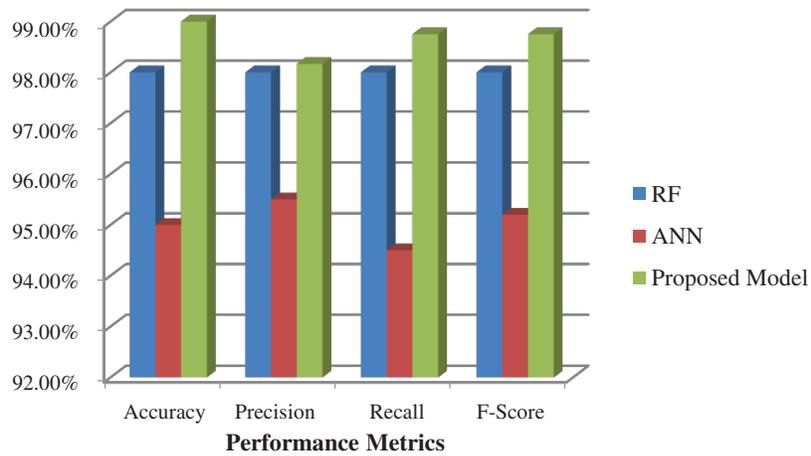

**Figure 8:** Performance metrics of the integrated model using N_BaIoT dataset

Fig. 9 shows the various performance metrics obtained using the SVM, NB, and DT classifiers on the BoT_IoT dataset. NB performs poorly with accuracy, precision, recall and f-scores of 87, 88, 88, and 87, respectively. The SVM performs better and results in an accuracy of 95%. The DT classifiers returned the best results with maximum accuracy of 98%. The performance of the proposed model after deploying the second level of classification using ANN and RF is shown in Fig. 10. The results of the merged prediction show an increase of 99% accuracy.

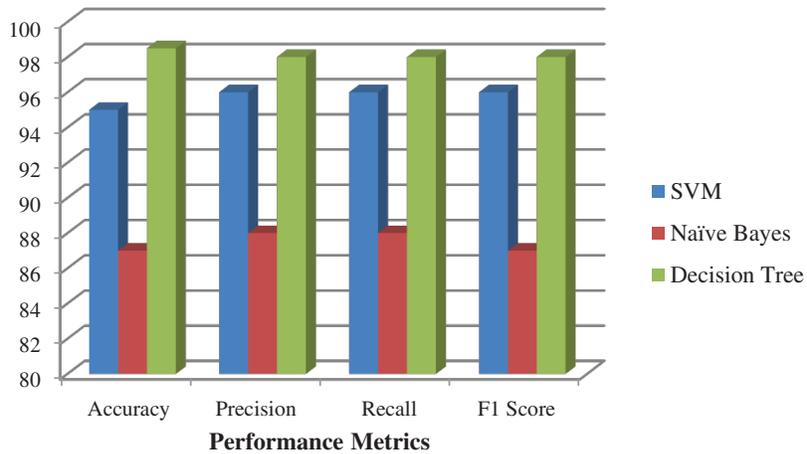

**Figure 9:** Performance metrics obtained on first level classification on BoT_IoT dataset

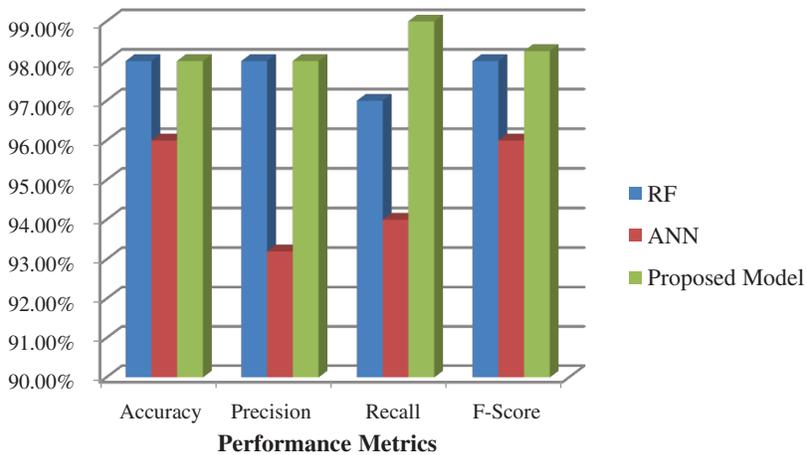

**Figure 10:** Performance metrics of the integrated model on BoT_IoT dataset

The major findings of the proposed work are as follows.

- Maximum accuracy of 99% is obtained for all three benchmark IoT intrusion detection datasets.
- The Adam optimizer increases the accuracy of the ANN and results in the overall best performance.

Tab. 4 lists the accuracy of the developed IIoT attack identification model compared to state-of-the-art intrusion detection models using multiple classifiers on the BoT_IoT dataset.

Table 4: Proposed model accuracy comparison with contemporary approaches

| Technique | Accuracy (%) |
| --- | --- |
| Stacking Ensemble of SVM and DT [35] | 94 |
| DeepDCA [30] | 98.7 |
| CNN [36] | 91.2 |
| Back-end LSTM [37] | 94.3 |
| Proposed IIoT Attack Identification Model | 99.7 |

## 5 Conclusion

Intrusion detection models are powerful mechanisms to secure IIoT systems. We conducted a literature survey of studies that investigated machine learning techniques on standard datasets to identify cyber threats and deployed identified learning approaches in our proposed model. The proposed model integrates three base classifiers, NB, SVM, and KNN by blending, i.e., a stacked ensemble technique. The second level classifier used in the proposed model is RF, and it is one of the best approaches to achieve higher prediction. The ANN and RF classification results are compared, and the best accuracy is considered the final result. The proposed model is evaluated on the WUSTL_IIOT-2018, N_BaIoT, and Bot_IoT datasets. Maximum accuracy of 99% with a marginal change in decimal values is obtained for all three datasets. Precision, recall, and F-Score values were also greater than 98%.

**Acknowledgement:** The authors extend their appreciation to King Saud University for funding this work through Researchers supporting project number (RSP-2020/164), King Saud University, Riyadh, Saudi Arabia.

**Funding Statement:** The authors extend their appreciation to King Saud University for funding this work through Researchers Supporting Project number (RSP-2020/164), King Saud University, Riyadh, Saudi Arabia.

**Conflict of Interest:** The authors declare that they have no conflicts of interest to report regarding the present study.